 \documentclass[preprint]{aastex}

\newcommand\etal{{\rm et~al\/}.} 
\newcommand\eg{{\rm e.g.\/}} 
 
\def\alwaysmath#1{\ifmmode {#1}  
                  \else {$#1\mkern-5mu$} \fi} 
\newcommand\km{\alwaysmath{\,{\rm  km}}} 
 
\newcommand\scd{\alwaysmath{\,{\rm  sec}}} 
\newcommand\kms{\alwaysmath{\,\km\,{\rm s}^{-1}}} 
 
\newcommand\angstrom{\alwaysmath{\,{\rm \AA}}} 
\newcommand\angs{\angstrom}

\newcommand\hr{\alwaysmath{\,{\rm  hr}}} 
\newcommand\mn{\alwaysmath{\,{\rm  min}}}

\newcommand\kpc{\alwaysmath{\,{\rm  kpc}}} 
 
\newcommand\kel{\alwaysmath{\,{\rm  K}}}
 
\newcommand\mg{\alwaysmath{\,{\rm mag}}} 
\newcommand\dg{\alwaysmath{\,^\circ}} 

\newcommand\astron{Astronomy}
\newcommand\dept{Dept.~}

\newcommand\das{\dept of \astron}

\newcommand\uof{University of~}

\newcommand\osu{Ohio State University}

\newcommand\uc{\uof California}

\newcommand\usyd{\uof Sydney}

\newcommand\aat{Anglo-Australian Telescope}
\newcommand\aura{Associated Universities for Research in Astronomy, Inc.}

\newcommand\ctio{Cerro Tololo Interamerican Observatory}

\newcommand\lick{UCO/Lick Observatory}

\newcommand\noao{National Optical Astronomy Observatories}

\newcommand\nsf{National Science Foundation}

\newcommand\ackaat{Based on observations obtained at the \aat}
\newcommand\ackaura{operated by the \aura} 
\newcommand\ackctio{Based on observations obtained at \ctio, which is \ackaura, under contract with the \nsf}

\newcommand\arw{
   \ctio, \noao, P.O. Box 26732, Tucson, AZ  85726; awalker@noao.edu}
\newcommand\dmt{
   \das, \osu, Columbus, OH 43210-1106; terndrup@astronomy.ohio-state.edu}
\newcommand\ems{
   School of Physics, \usyd, Sydney, NSW 2006 Australia; ems@physics.usyd.edu.au}
\newcommand\rcp{
   \lick, \das, \uc, Santa Cruz, CA 95064; peterson@ucolick.org}
\newcommand\ilc{i}
\newcommand\iilc{ii}

\newcommand\logg{\alwaysmath{ \log\,g }} 
 
\newcommand\teff{\alwaysmath{T_{{\rm eff}}}}

\def\popi{\alwaysmath{\rm Population\,{\textsc \ilc}}}
\def\popj{\alwaysmath{\rm Population\,{\textsc \iilc}}}

\newcommand\feh{{\rm [Fe/H]}}
\newcommand\alfe{\alwaysmath{\rm [\alpha/Fe]}}
\def\2h#1{\alwaysmath{{\rm [#1/H]}}}
\def\2fe#1{\alwaysmath{{\rm [#1/Fe]}}}

\newcommand\caj{{\rm Ca\,{\textsc \iilc}}}

\newcommand\srj{\alwaysmath{\rm Sr\,{\textsc \iilc}}}
   
\newcommand\ebmv{\alwaysmath{E(B-V)}}
 
\newcommand\bmv{\alwaysmath{B-V}} 

\newcommand\umb{\alwaysmath{U-B}} 
 
\newcommand\vmi{\alwaysmath{V-I}}

\newcommand\dmv{\alwaysmath{(m-M)_V}}
 
\newcommand\mv{\alwaysmath{M_{_{V}}}}

\newcommand\Halpha{\alwaysmath{{\rm H}\alpha}} 
\newcommand\Hbeta{\alwaysmath{{\rm H}\beta}} 
\newcommand\Hgamma{\alwaysmath{{\rm H}\gamma}} 
\newcommand\Hdelta{\alwaysmath{{\rm H}\delta}} 
 
\newcommand\vt{\alwaysmath{\chi_{\rm t}}} 
\newcommand\vsini{\alwaysmath{v \sin \ilc}} 
 
\newcommand\ngc{\alwaysmath{\rm NGC\,}}

\slugcomment{To appear in {\it The Astrophysical Journal} v547n1 20 Jan 2000}

\shorttitle{Bulge BHB Stars}
\shortauthors{Peterson \etal}

\begin{document}

\title{Hot Horizontal Branch Stars in the Galactic Bulge. I.\altaffilmark{1,2}}

\author{
Ruth C. Peterson\altaffilmark{3}, 
Donald M. Terndrup\altaffilmark{4}, 
Elaine M. Sadler\altaffilmark{5}, and 
Alistair R. Walker\altaffilmark{6}
}

\altaffiltext{1}{\ackaat.}

\altaffiltext{2}{\ackctio.}

\altaffiltext{3}{\rcp.}

\altaffiltext{4}{\dmt.}

\altaffiltext{5}{\ems.}

\altaffiltext{6}{\arw.}

\begin{abstract}

We present the first results of a survey of blue horizontal branch
(BHB) stars in the Galactic bulge. In this exploratory study,
candidates with $15 \leq V \leq 17.5$ covering a wide range in
\bmv\ color were selected from CTIO Schmidt $UBV$ photometry. Blue
spectra were recorded at 2.4\angs\ FWHM resolution for 164 stars in a
1.3 sq.\ dg.\ field $\sim$7.5\dg\ from the Galactic center. Radial
velocities were measured for all stars.  For stars with strong Balmer
lines, we devised and applied a spectroscopic technique to determine
stellar temperature \teff, gravity \logg, and metallicity
\feh\ independent of reddening. The reddening and distance to each star were then found from $UBV$ photometry.  Reddening proved highly
variable, with \ebmv\ ranging from 0.0 to 0.55 around a mean of 0.28.
The \bmv\ colors of cool HB stars of solar metallicity reddenened by 
\ebmv\ $\geq 0.3$
overlap those of foreground main-sequence stars, but the \umb\ vs.\ \bmv\ 
diagram distinguishes these groups until \ebmv\ $>$ 0.5.  

Forty-seven BHB candidates were identified with \teff\ $\geq$
7250\kel.  Seven have the gravities of \popi\ stars, three are
ambiguous, and 37 are HB stars, including perhaps a dozen RR
Lyraes. The unambiguous BHB stars are all cooler than 9000\,K. They span a wide
metallicity range, from solar to 1/300 solar.  The warmer BHB's 
are more metal-poor and loosely concentrated towards the Galactic center,
while the cooler ones are of somewhat higher metallicity and are situated closer 
to the center. We detect two cool solar-metallicity HB 
stars in the bulge of our own Galaxy, the first such stars known. Still elusive 
are their fainter hot counterparts, the metal-rich sdB/O stars strong in ultraviolet light.

\end{abstract}


\section{Introduction}

According to classical calculations of single-star evolution
\citep[e.g.,][]{roo73}, the color of a star that has left the giant
branch to become a core-helium-burning star on the horizontal branch
(HB) depends primarily on its age and metallicity. At lower
metallicities, a wide range in color is both found and predicted among
BHB stars if modest mass loss is assumed.  
Among populations of solar metallicity or higher, only HB
stars redder than the RR Lyrae instability strip should be produced
within a Hubble time.  However, the most metal-rich globular clusters
in the Galaxy, such as \ngc 6388 and \ngc 6441 near the Galactic
center, do show a handful of BHB stars \citep{ric97}. These are mostly
cool or warm BHB's \citep{moesc99}, with few if any of the hottest, faintest 
types, the subdwarf sdB stars and their relatively rare successors the sdO
stars.

Surprisingly, significant numbers of sdB/O's are found in metal-rich
populations in the Milky Way; even more surprisingly, 
they appear to outnumber cool and warm metal-rich BHB's.  In the
field, where it is difficult to distinguish BHB's of intermediate
temperature from \popi\ A stars, few if any metal-rich cool BHB's are known --
\citet{gra96}, for example, uncovered only one possible BHB with \feh\ $\ge$ $-0.7$,
out of 39 BHB's newly identified. In contrast, 
there are dozens of field sdB/O stars whose gravities indicate HB status, and
whose radial velocities indicate a thick disk rather than halo population
\citep{saf94,saf97}. Among open clusters \citep[reviewed by][]{fri95}, 
the most striking occurrence of sdB/O stars is found in
\ngc 6791, with a metallicity 3 -- 4 times solar
\citep{pet98}. In addition to two cooler BHB (or blue straggler)
members, the cluster harbors four or five sdB/O stars \citep{lie94}
whose membership is likely given their spatial concentration towards
the cluster center \citep{kal92}.  

Hot, metal-rich BHB stars also appear to be present in metal-rich extragalactic
systems.
As reviewed by \citet{oco99}, elliptical galaxies and early-type spiral
bulges commonly show an upturn in integrated light below 2000\angs,
where old main-sequence turnoff stars have very little flux. Both the
smooth spatial distribution and the continuous spectral distribution of
the UV upturn indicate that it is not caused by young O and B stars,
but rather by hot old stars, sdB/O's and the brighter but rarer blue
post-AGB stars \citep{dor95,bro97}. 
Among elliptical galaxies, the size of the UV upturn
tends to increase with increasing galactic metallicity
\citep{fab83,bur88,lon89}, although the strongest correlation is with
line indices based on light elements rather than iron itself, the scatter is large, and there is no continuity between these systems and globular clusters. 

The presence of BHB stars in high-metallicity populations and the reversal 
in their color distribution raise major questions as to their origin.  
Both observation and theory suggest that
BHB stars in metal-rich systems may be produced by channels in addition
to those operating in metal-poor systems. 
Theoretical production of metal-rich sdB's from single-star evolution can be 
achieved with a rapid increase at  \feh\ $> 0$ in either helium abundance 
\citep{bre94,yi98} or mass loss \citep{dor95,dcr96}. It may also be 
accomplished with deep mixing, and thus accompanied by light-element 
enhancements \citep{swe97,kra98}. However, binary mass transfer may play a 
dominant role at the hot end of the BHB, since a large fraction of field 
sdB/O stars are found to be binaries \citep{all94,gre00}. Blue stragglers, thought 
also to have formed by mass transfer, may therefore be involved in the 
production of hot HB stars. 

Constraints on these mechanisms might be placed 
by determining the color distribution and light-element ratios 
of metal-rich versus metal-poor BHB's, and by determining 
whether blue stragglers of similar metallicity are invariably present.
The best place to do this is in the Milky Way bulge
itself. The bulge is the only Milky Way population sizable enough to
support large numbers of BHB stars. Unlike the centers of external
galaxies, it is sufficiently nearby that stars as faint as the sdB's
may be resolved individually. It resembles elliptical galaxies and
spiral bulges in stellar density and star-formation history
\citep{whi78,fro87,ter90,hou95} as well as in a relatively high
abundance of light elements \citep{mcw94,sad96}. Its metallicity
gradient \citep{ter88,fro90,tys93,tie95} provides a natural
testbed of how metallicity drives stellar evolution.

Consequently we have undertaken a survey of the BHB population in four
windows of the bulge along and near its minor axis. 
BHB candidates are selected from $UBV$ photometry. 
For a representative subset, 
follow-up moderate-resolution spectroscopy is providing 
stellar parameters (confirming BHB status and establishing reddening), 
plus iron and magnesium abundances to as high a temperature as possible. 
The basic goal is a
statistically complete survey of stars on or near the BHB across all 
temperatures where they occur, from 7250\kel\ to 35,000\kel. We aim 
for the same degree of completeness for all BHB stars, regardless of 
their temperature and metallicity.
Such a survey would show immediately whether cool BHB's
and sdB/O stars exist at all in the bulge, as suggested at the cool end
from its RR Lyrae stars \citep{wt91}. 
Knowledge of the numbers of hot BHB stars in each field, in
conjunction with the metallicity and temperature distribution of cooler
BHB stars in the same region, should help greatly in disentangling the
relative influence of the various production factors noted above.

We begin with this pilot study of a single region
limited in magnitude to the cool end of the BHB, 
whose results we briefly summarize here.
Spectra were obtained during the commissioning phase of the
Two-Degree Field (2DF) spectrograph of the \aat. Forty-seven stars whose
Balmer-line profiles indicated temperatures $\geq$ 7250\kel\ were
analyzed by comparing their spectra with a grid of theoretical
spectra.  Reddening was then found from the model colors as tabulated by Kurucz.
We show that reddening varies dramatically from star to star within this field,
with $0 \leq\ \ebmv\ \leq\ 0.55$,
so that colors alone cannot determine \teff\ and \logg.
However, the $UBV$ color-color diagram helps to distinguish BHB stars 
from main-sequence turnoff interlopers as long as \ebmv\ $\la$ 0.50. 

Thirty-seven of the hot stars proved to have temperatures and gravities
indicating a position on the horizontal branch. 
Two more of the hottest stars might be either BHB or main-sequence stars.
None of the unambiguous BHB stars in this sample proves to have \teff\ 
$\geq$ 9000\,K, which we attribute to the sparseness of our sample at blue colors 
and faint magnitudes. We estimate roughly a dozen 
of our coolest hot stars to be RR Lyraes, pulsating variables
located just redward of the BHB stars on the horizontal branch.
The hotter BHB stars in this sample tend to be more metal-poor and more spatially
extended than the cooler ones. We comment on the possible implications 
for BHB and RR Lyrae production in metal-rich and metal-poor bulge populations.
Five HB stars are discerned with \feh\ $\geq$ $-0.5$ in the bulge itself,
the first such stars found. Searches to fainter magnitudes are planned to reach
the hot sdB stars believed responsible for the UV upturn.

\section{Photometric Observations and Data Reduction} 

During a seven-night run in 1995 May-June, we used the Curtis Schmidt Telescope 
at \ctio\ (CTIO) to obtain $UBV$ images of a total of 11 fields of the bulge: 
five overlapping fields between $-6$\dg\ and $-8$\dg\ along the minor axis, and 
three displaced 3\dg\ on each side of $-6$\dg. The field discussed here was 
observed on 1995 May 31 UT, located at $(\ell,b) = (-3.325, -6.731)$, with 
central coordinates $\alpha = 18^{\rm h} 05^{\rm
m} 17^{\rm s}, \delta = -35^\circ 10^{\prime} 51.00^{\prime\prime}$
(2000.0).  We will call this our ``target'' field, as compared to a
calibration field discussed below.  

The CCD detector provided a scale of 2.024\arcsec\,pixel$^{-1}$ in $V$, for a 
field of view of 1.15\dg\ on a side, or $1.3$ sq.\ deg.\ The exposures were taken 
in the sequence $8 \times 90$ sec in $V$, $8 \times 300$ sec in $B$, and $10 
\times 300$ sec in $U$. The limiting magnitudes were about $18.7$, $19.2$, and 
$18.2$ for $V$, $B$, and $U$  respectively; about 60,000 stars
per square degree were detected with both $B$ and $V$ photometry.

The basic processing of the frames in the target field consisted of
overscan and zero corrections, and division by median twilight sky
flats.  This step was performed using the {\tt ccdred} package of the IRAF
reduction package, available at http://iraf.noao.edu/.

Stellar positions and magnitudes were measured on each individual CCD
frame using DAOPHOT II \citep{stet87}.  Initial positions were
determined by finding stars on a combined frame in each filter, where
the individual images were shifted to a common coordinate system and
then combined using a sigma clipping algorithm to reduce cosmic rays.
The photometry, however, was performed on the uninterpolated images,
as the stellar profiles were undersampled.  The PSFs, which showed
significant variation with position, were determined for each frame
using between 50 and 200 bright, uncrowded stars;  the positions of the
PSF stars were checked to insure that they were uniformly distributed
over each frame.  The PSF was modeled to have FWHMs in $x$ and $y$
which varied quadratically with position in the frame.  The relative
magnitudes were determined with the ALLSTAR routine in DAOPHOT II in a
single pass (i.e., no additional stars were subsequently found on
PSF-subtracted images and added to the list of trial positions).

The resulting lists of magnitudes and positions were then transformed
to a common coordinate system, defined by the positions on one of the
$V$ frames.  The correction included terms for pixel scale (identical
in $x$ and $y$) and for rotation;  the latter correction, however, was 
insignificant.

The next step was to combine the several magnitudes in each filter into
three lists of average instrumental magnitudes in $UBV$.  The method
was to pick one of the exposures in each filter as a standard, then to
determine a single offset to bring the magnitude scales into
agreement.  These magnitude offsets showed small but significant
variations, indicating that the transparency varied during the $UBV$
series.  When the magnitude offsets were determined for all the frames
in one filter, these offsets were applied and the resulting photometry
was averaged.  Specifically, the weighted average and weighted standard
deviation were computed for each star, where the weights were
determined from the errors in a single measurement reported by
DAOPHOT.  The standard deviation of the (x,y) positions gave a measure
of the positional accuracy, which was about 1/30 pixel for the bright
stars and decreasing to about 1/3 pixel at the detection limit.  Stars
which were detected in at least three frames on each filter were
kept and the others discarded.

Photometric calibration was based on a series of $UBV$ images 
obtained for an overlapping field on the photometric night
of 1995 June 1 UT.  These calibration frames were offset from the
target field by $7\arcmin$ in right ascension and $2\arcmin$ in
declination, in order to calibrate several Schmidt fields at once.
The exposure sequence was $2 \times 90$ sec
in $U$,  $2 \times 60$ sec in $B$, and $2 \times 30$ sec in $V$.  
Basic image processing, the extraction of instrumental
magnitudes and averaging of the photometry proceeded as for the target
field. On the same night, we also observed many fields with E-region
standards in \citep{gra82}, obtaining aperture photometry of the
standards and computing transformation equations in $V$, $B-V$, and
$U-B$ including airmass corrections.  We then applied these
transformations and computed the zero points to bring the
deeper photometry from the target field onto the same scale.

An astrometric transformation was obtained from the $V$ frames 
by using the search engine at
http://www.cfht.hawaii.edu/$^\sim$bernt/gsc/gscbrowser.html
to identify 271 guide
stars from Hubble Space Telescope.
An astrometric solution was
found which included a single frame scale in $\alpha,\delta$, a
rotation angle, and quadratic terms in row and column position.  The
resulting solution had an r.m.s. scatter of $0.4\arcsec$ in right
ascension and declination (i.e., about 1/5 pixel), adequate for our
multifiber spectroscopy.

Figure 1 shows the color magnitude diagram (CMD) in $V$, $B-V$
for our target field. Stars observed spectroscopically are indicated as large  
symbols, with filled circles representing those stars found to be 7250\kel\ 
or hotter in the analysis described below. Small dots mark the positions of 
17\% of the remaining stars. The same designations are 
maintained in Figure 2, a color-color diagram plotting \umb\ vs.\ \bmv\ 
in the vicinity of the main-sequence turnoff of foreground stars. 
As discussed in \S 6, this diagram helps discriminate lightly-reddened 
foreground stars from hotter but more heavily reddened stars, 
including the HB stars of interest.

\section {Spectroscopic Observations and Data Reduction}

In this exploratory work, our selection of candidates was based on the CMD only, 
to ensure a completely unbiassed sample. 
From well-populated regions of the CMD, we drew stars 
broadly distributed in color as targets for fiber spectroscopy.
The magnitude and color limits, $14.8 \leq V \leq 17.8$ and $B - V \leq 1.2$, 
include HB stars within and beyond the bulge over a wide range of metallicity and 
reddening. Fiber tangling and minimum proximity ($\geq$ $11.5\arcsec$ in this case)
ruled out some candidates once higher-priority targets are selected; we 
gave higher priority to bluer stars. 

On 1997 July 3 UT, the Two-Degree Field (2dF) 
instrument at the \aat\ \citep{lgt98,sl98} was used to obtain $6 \times 
1800$\,sec exposures with a single fiber configuration on the target field.  188 
fibers $2.16$\arcsec\ in diameter were placed within its 2\dg\ field of view.
A total of 164 fibers were 
assigned to stellar targets, and 24 assigned to sky regions 
chosen to be free of resolved stars by inspection of the $B$
and $V$ CCD frames.  
The 1200B grating in the standard AAT set was employed in first order,
yielding a resolution of 1.11\angs\,pixel$^{-1}$ and a wavelength
coverage of 3834 -- 4973\angs.  The effective resolution, judged from
the width of the emission lines in the terrestrial sky, is about 2.0
pixels, or 150\kms\ at the central wavelength of the spectra.  Spectral
synthesis indicates 167\kms\ or 2.2 pixels for the coadded target
observations.

The spectra were extracted using the 2DFDR reduction pipeline described
by \cite{bg99}.  Cosmic rays were removed
from the individual exposures using a sigma-clipping algorithm, and
the cleaned images were summed to produce an image with an
effective exposure time of 10,800\scd.  The wavelength scale was
determined from emission lines in a CuAr spectrum.  The extracted
spectra were scaled to correct for the nonuniform fiber efficiencies.
The sky spectra were then averaged together and subtracted from each of
the object spectra.

The mean counts per pixel in the extracted spectra correlated tightly
with magnitude, from about 8500 counts at $V = 15.5$ to 2000 counts at
$V = 17.5$, verifying the accuracy of the astrometry.  The brightest
and faintest spectra have signal to noise of $\sim$ 125 and 30 per
2.2-pixel resolution element at 4800\angs, decreasing to about half
this at 3900\angs\ for the hot stars.

\section{Radial Velocities}

To measure radial velocities, the spectra were rebinned to a log-wavelength
scale and inspected by eye.  They fell into two distinct classes ---
F-star spectra with narrow absorption lines of hydrogen and
metals (94 stars), and BHB-like spectra dominated by strong, broad Balmer
absorption lines (38 stars). 15 spectra were classified as intermediate
between these two classes, and about 20 spectra had low S/N and were omitted
from further analysis.

The velocities were measured from the {\tt scross} cross-correlation routine
of the Figaro reduction package \citep{sho95}. 
We constructed two radial-velocity templates by co-adding spectra with
good S/N.  One was the sum of 42 F-star spectra, the other 
the sum of 11 BHB-like spectra.  The radial velocity for each star was found
by cross-correlation with the appropriate template; for the ``intermediate''
stars, we used both templates and took the average value. Velocities are
relative to the F-star template, and have a random error of 5 -- 10\kms, 
depending on exposure level and line strength. This estimate is based on the 
agreement with velocities determined by visual comparison with synthetic spectra. Most stars' observed spectra matched well after 10\kms\ was subtracted to bring the tabulated velocities onto a geocentric scale. For stars 168, 75, 110, 118, 190, and 57, an additional 10\kms\ needed to be subtracted from the tabulated velocities to match the synthesis; for stars 126 and 192, 10\kms\ had to be added back; for stars 52 and 66, an additional 30\kms\ had to be subtracted.

The velocity dispersions of the F-star and BHB-star groups are significantly
different. (The ``intermediate'' stars had a velocity distribution similar to
the BHB stars, and so they were added to the BHB group).  The 94 stars with
F-star spectra have a mean velocity of $-3.9$\kms\ and a dispersion of
51.0\kms, while the 53 BHB and intermediate stars have a mean velocity
of $-2.9$\kms\ and a dispersion of 120.9\kms.
There is no obvious trend of radial velocity with apparent magnitude, and the
mean magnitude of the two groups is very similar.
On the basis of their kinematics, then, we identify the F-star group as
members of the foregound Galactic disk, for which we expect a velocity
dispersion of 40--95\kms\ depending on their mean distance \citep[\eg,][]{ter95},
and the BHB group as primarily bulge stars. 

\section{Spectrum Analysis}

Spectrum analysis was based on visually matching each observed spectrum
to a sequence of {\it ab initio} calculations based on model
atmospheres and a line parameter list. The program used, SYNTHE
\citep{kur81}, was downloaded from the Kurucz Web site at
http://cfaku5.harvard.edu, and modified by S. Allen and B. Dorman to
run on an Ultra-30 at the University of Virginia. We also downloaded
the \citet{kur91} grid of ATLAS9 models and colors, and the line lists
$gf$0400.100 and $gf$0050.100 dated 25-May-98. As described by
\citet{kur95}, they differ from those of \citet{kurbel95} in
incorporating for many iron-group species all newly identified and
revised transitions whose energy levels were measured recently by
\citet{nav94} and other laboratory groups. Moreover, they include
only atomic lines identified in the laboratory. 
The predicted lines are of no use here because their wavelengths are 
good to only $\sim$10\angs\ (being generated by Kurucz from his 
semi-empirical atomic models, in which 
laboratory-based energy levels were extrapolated to high excitation). Their 
omission has no effect on derived parameters, since very few unidentified lines 
appear in these optical stellar spectra, which are generally weaker-lined 
than the Sun. 

Model metallicities ranged from 1/300 solar to twice solar: \feh\ =
$-2.5$, $-2.0$, $-1.5$, $-1.0$, $-0.5$, 0.0, and $+0.3$.  The
light-element abundance ratios were held fixed at \alfe\ = 0.0 (solar)
for \feh\ $\geq$ $-0.5$, \alfe\ = +0.2 for $-1.0$ and $-1.5$, and
\alfe\ = +0.4 for \feh\ $\leq$ $-2.0$, with O, Mg, Si, Ca, and Ti
abundances so adjusted. This was done because most of the light-element
lines are blended with iron-peak lines at this resolution; at higher 
resolution the \alfe\ ratio could be determined independently.
In all cases, the iron abundance was reduced by
0.15 dex to the currently accepted solar value, and the model
ionization equilibria were recalculated.

Each stellar spectrum was calculated from 3555 to 4965\angs\ at a
resolution of 300,000, and broadened by a macroturbulent velocity $v_m$
= 1.5\kms, a microturbulent velocity \vt\ = 2.0\kms, and a Gaussian
instrumental profile of FWHM 167\kms\ (2.4\angs\ at 4400\,\AA)
which best matched the observed spectra. For two stars noted in \S 8, the observed
spectral features required additional broadening.

With few exceptions, we analyzed only those stars with temperature
\teff\ $\ge$ 7250\kel, that of the metal-rich star 2-17 
situated near the cool end of the BHB in \ngc 6791,
either a BHB or a blue straggler \citep{pet98}. 
Our spectral calculations showed that at the resolution
of these data, the central flux of \Hbeta\ is 39\% that of the
continuum for a cool BHB star at 7250\kel\ with \logg\ = 2.5. To be
sure to identify all stars with \teff\ $\geq$ 7250\kel\ in the presence
of noise or a binary companion, we accepted for further consideration
any spectrum with a relative central \Hbeta\ flux of 52\% or less. One
star with broad but shallow Balmer lines was also included. We had
tried using more standard criteria based on the breadth of the
\Hbeta\ wings at 80\% (or 70\%) flux level and the total
\Hbeta\ equivalent width, but both approaches failed at this resolution
and with this wide range of metallicity, because atomic-line blends in
solar-metallicity stars mimic moderately stronger Balmer wings and
generate larger Balmer equivalent widths.  As illustrated below, this blending is worse for  \Hgamma\ and prohibitive for the other Balmer lines.

Because of this blending, setting the stellar temperature \teff\ was
done using the central $\sim$5--8\angs\ of the profiles of \Hbeta\ and
\Hgamma, excluding as noted below the central 1--2\angs\ of the core 
(with the larger
values referring to hotter stars). This approach is totally insensitive
to reddening, and we found it to be independent of \feh\ as well and
largely insensitive to \logg. An initial model was chosen to
approximately match both Balmer and metal lines; its spectrum was
calculated and shifted by the stellar velocity, then compared to the
observed spectrum. A new model was selected to address any mismatch
(with different \feh\ and velocity if necessary) and the procedure
repeated until the agreement was satisfactory. A few cases never
reached this goal; as discussed below, most have anomalous Balmer line
profiles and are likely to be RR Lyrae stars. An alternative possibility
of a high helium abundance (with helium and hydrogen equal by number) was
proposed by \citet{gra96} for two field BHB stars 
that they found to 
have Balmer lines too broad for the Balmer jump.

Once \teff\ was set, the breadth of the Balmer line wings was used to
determine the gravity, by fitting the wings of the \Hbeta\ and
\Hgamma\ profiles. Final adjustment was made to \feh\ to best reproduce
the strengths of the atomic lines, again as judged by eye.
We analyzed fully a total of 60 stars, including four, seven, and two
with \teff\ = 7000\,K, 6750\,K, and 6500\,K respectively as
templates to weed out the other cooler spectra.  

Reddening was derived for each star with $UBV$ photometry.  The value
of \ebmv\ to the nearest 0.05\mg\ which best reproduced the \umb\ and
\bmv\ colors was found from the tabulation by \citet{kur91} of $UBV$
colors vs.\ \ebmv.  For each star whose parameters place it
near either the BHB or the main sequence, an \mv\ value was assigned accordingly
and a distance then determined from the observed photometry and deduced
\ebmv.

Table 1 lists basic observational data and derived stellar parameters
for each of the 47 stars found to have \teff\ $\geq$ 7250\kel. The first
column gives the 2dF fiber number, by which we subsequently refer
to individual objects. The next columns list quantities determined
during the photometric and astrometric reductions: the star ID, position, $V$
magnitude, \bmv\ color, and \umb\ color and their uncertainties, the 
1$\sigma$ deviation of the measurement from an individual frame about the mean.
Next appears the radial velocity V$_{\rm rad}$ in \kms\ on
the internal 2dF scale. The following columns give the deduced \teff, \logg,
and \feh\ (designated $Z$), and the final columns give parameters
deduced by fitting the colors of the model at the reddening listed to those observed
for the star. D is the distance from the Sun in \kpc; \mv, the absolute
visual magnitude; $E$, the reddening \ebmv; and $\delta_U$ and $\delta_B$,
the difference between the observed and the model \umb\ and \bmv\ colors.

In Figure 3, we show the fits achieved for stars spanning a wide range
in parameters and in S/N. Each panel compares the observed and calculated spectra
for the same six stars in a different wavelength region.
In each comparison, the heavy line is the observation and the light line the 
calculation; successive spectra are displaced upwards by 20\%. 
The model atmosphere parameters 
are listed on the right with the fiber number of the star
observed. 

We note immediately 
that the very center of each Balmer line is always too deep in the 
calculated spectra. We have verified that the core and only the core of the 
Balmer lines is mismatched in BHB stars, using echelle spectra of field BHB stars. 
For example, for HD 130095, we match the \Hgamma\ line profile to 2\% or better 
until +/- 1.2A of the center, where the residual flux has dropped 
below 30\% of the continuum (Peterson et al., in preparation). In the Sun, 
the core of \Halpha\ is formed well into the chromosphere, where CaII emission 
arises \citep{dup92}. We thus attribute this core 
mismatch to uncertainties in the superficial layers of cool stars with convection 
and chromospheres, and ignore the line core in determining stellar 
parameters.

In addition to the strong Balmer lines, numerous weak 
lines of heavier elements appear in Figure 3. 
Their strength and numbers generally increase 
towards blue wavelengths and in cooler stars. 
Thus the Balmer-line profiles, especially \Hdelta, are seen to be contaminated
by weak atomic lines in the cool, metal-rich stars, precluding the 
use of standard measures of Balmer-line breadth to derive \teff\ and \logg.
That our own procedure succeeds is indicated by the goodness of fit of the spectra in all panels of Fig.\ 3, in particular \Hdelta\ in panel a.

To explore uniqueness of the determinations,
alternative fits were tried. In Figure 4, a second calculation 
is shown for the \Hgamma\ and \Hbeta\ regions 
of the spectrum of each star of Figure 3. 
The same model parameters were assumed except that surface gravity 
was taken to be \logg\ = 3.5 (4.0 in the bottom spectrum). 
Not shown is a third calculation 
in which the same gravities were adopted as in Figure 4, but \teff\ was changed by 250\kel\ in an attempt to compensate for the change in line wings introduced by the change in \logg. The resulting fits to \Hgamma\ and \Hbeta\ were the same or worse as in Fig.\ 4.

Comparing Figures 3b and 4a, and Figures 3d and 4b, indicates 
that gravities are well-determined for the hotter stars, 
but by 8000\kel, the sensitivity of \Hgamma\ and \Hbeta\ to
a change of 0.5 dex in \logg\ has become minimal. 
At lower temperatures, the weak Balmer profiles make the gravity determination difficult at high metallicity and low S/N. 
The top three spectra illustrate the increasing difficulty of establishing \logg\ at \teff\ $\leq$ 7500\kel.

Nonetheless, the best fits are clearly 
better in both Balmer lines than the alternative fits, in all but the top 
spectrum. Even there, the best fit is marginally better. Moreover, 
blueward of 4315\angs\ in the top spectrum only, 
the band head of the CH molecule is evident as the absorption unmatched by the spectral
synthesis. Because molecular lines become strong only in cool, metal-rich,
and/or high surface-gravity stars such as this one, they were not
included in the spectral synthesis line list. The appearance of CH here
supports the low \teff\ and high \logg\ we deduce for this star. 

We also ran comparisons to test the effect of the treatment of convection. \citet{cas97} have shown that turning off convective overshoot in Kurucz models changes the temperature of deep layers in cool stars. Colors and Balmer-line profiles are affected, especially at low metallicities and in late A stars where convective transport is strongest. \citet{pet00} find that Castelli et al.\ models but not \citet{kur95} models simultaneously match the \Halpha\ profile and the mid-UV flux distribution in metal-poor turnoff stars with \teff\ $<$ 7000\kel. We thus calculated additional spectra for stars 57 and 98 using Castelli et al.\ models (downloaded from the Kurucz web site). At \teff\ = 7500\kel, \logg\ = 3.0, and solar metallicity, no change in the \Hbeta\ line profile was seen. At \teff\ = 7250\kel\ and \feh\ = $-1.0$, the wings of \Hbeta\ became too strong to match star 57 at any \logg. At \teff\ = 7000\kel\ and \logg\ = 3.0 a good fit was found for \Hbeta, but the wings of \Hdelta\ were too weak. Unfortunately, our photometry shows that this star is variable (\S 7). Further tests with higher-quality data for nonvariable stars are needed to elucidate the effects of the treatment of convection on Balmer-line profiles in cool BHB stars.

From comparisons such as this, we estimate uncertainties as follows. 
The uncertainty in \teff\ for the stars with \teff\ $\geq$ 7500\kel\ is
250\kel, and in \logg\ is 0.5\,dex, provided \Hdelta\ is also matched in stars with \teff\ $\leq$ 8000\kel. For metal-poor stars of 7250\kel, the same errors apply in gravity, to which the Balmer-line wings are still somewhat sensitive, but the treatment of convection may increase errors in \teff. For stars of 7250\kel\ with near-solar metallicity, evolved stars can be distinguished from main-sequence stars via the CH band. 

An additional check on the validity of the deduced atmospheric parameters was made
from the residuals in \bmv\ and \umb. For stars with
\teff\ $\ge$ 8000\kel, the 1$\sigma$ standard deviation of the
\bmv\ and \umb\ colors is marginally greater than the
average uncertainties of measurement.  
However, several cooler stars show significantly larger discrepancies and larger errors in stellar parameters.  These may be RR Lyrae variables.
We discuss these below, after examining the consequences of the large range in 
reddening.

\section{Reddening}

For the total
of 36 HB stars with $UBV$ photometry, $\langle\ebmv\rangle = 0.28$, and the
1$\sigma$ deviation about the mean is 0.14. 
However, reddening differs dramatically among the hot stars:  
values from \ebmv\ = 0.0 
to 0.55 occur throughout the 1.3 sq.\ deg.\ field.
A similar mean and range in \ebmv\ are found
among only those stars with \teff\ $\ge$ 8000\,K, almost all of which
are BHB stars, indicating that RR Lyrae variability 
is not to blame.  

With such a range in reddening, 
$UBV$ photometry by itself cannot be used to simultaneously determine 
\teff, \logg, and \feh\ of a star, since \bmv\ and \umb\ are both
sensitive to all four parameters. Many recent studies of BHB stars in the field
halo \citep[\eg,][]{kin94,wil99} simply assume a reddening value, usually
\ebmv\ $<$ 0.10. While this may be appropriate for the 
halo, it is clearly unwarranted here, and presumably elsewhere at
low galactic latitude. In contrast, fitting high S/N spectra of 
2.4\angs\ FWHM resolution to the Balmer lines \Hbeta, \Hgamma, and \Hdelta\ does provide
a reliable way to disentangle the parameters for stars with
7250\kel\ $\leq$ \teff\ $<$ 10,000\kel.  The method works because the
blends are explicitly modelled, the inner few \angs\ of the Balmer
lines outside the core generally reflects \teff, and the wings then give \logg; reddening follows from the colors.

When moderately reddened, the more metal-rich hot stars near the red end of the 
BHB have \bmv\ colors which overlap 
those of main-sequence foreground stars. This is evident in Figure 1,
where the hot stars (filled circles) are predominantly blueward of the 
main-sequence crush but include many stars within it. Even in the presence of 
moderate reddening, however, the \umb\ vs.\ \bmv\ color-color diagram aids in 
separating hot stars from solar-metallicity turnoff stars. 
This separation of hot and cool stars is 
illustrated in Figure 2. Empirically, the majority of the 
hot stars have a bluer \bmv\ color at a given \umb\ than the 
cooler ones (open circles). (Many of the most deviant filled circles are the 
probable variables listed below, and the open 
circles in the BHB region may be RR Lyraes which were cooler than 7250\kel\
at the time of the spectroscopic observations.) 
The theoretical model colors concur, as shown by the curves. They were generated from 
the tabulated colors of models with \feh\ = 0 by interpolating in \logg\ 
along the zero-age BHB, 
from \teff, \logg\ = 7250\kel, 2.75 to 13,000\kel, 4.1. The three curves represent 
reddenings of \ebmv\ = 0.0, 0.3 (the mean for the field), and 0.5; the 
reddening vector is given by the arrow, whose heads correspond to the latter 
reddenings. The figure illustrates that the color-color diagram provides discrimination 
against field main-sequence stars for BHB stars of all temperatures until \ebmv\ 
exceeds 0.5, when BHB's cooler than $\sim$ 15,000\kel\ are superimposed on them. 
At higher reddening, it might
be thought that $I$ band colors would help, but the model colors indicate this
is true only if the $I$ magnitude is measured to $\pm$0.05\,mag. $JHK$ 
photometry would also help in principle, but requires different detectors which currently
have a smaller field of view, and leads to worse crowding by redder stars.

\section{RR Lyrae Interlopers}

As noted in the review of RR Lyrae stars by \citet{smi95},
the pulsational cycle that drives RR Lyrae 
variability leads to temperatures of RR Lyrae stars near maximum light that 
overlap those of stable BHB stars. The spectral type at maximum of 
RR Lyraes is A7--A8, corresponding to $\sim$8000\kel. 
Thus a significant fraction of RR Lyraes is anticipated among our cooler HB 
sample.

One indication of their presence is the occasional large size of deviations 
in \umb\ and/or \bmv\ color 
seen for stars cooler than 8000\kel\ in Table 1. Of the hotter stars, only star 162
shows a deviation $>$ 0.12, but star 163 does so at 
7750\kel, and six cooler stars do as well.
Another indication comes from anomalous Balmer lines. 
Star 162 shows distinct Balmer-core emission, and an increasing fraction of
cooler stars have Balmer lines with asymmetric profiles or 
filled-in cores or very broad wings. The majority of
the stars with large photometric deviations show such anomalies. 

We suspect that these are signs of pulsational variability. 
Metal-poor RR Lyrae stars of
type $ab$ are known to show Balmer-line emission during the rise to
maximum light \citep{pre64}, due to shocks formed by pulsation; the
milder pulsation of the overtone $c$ variables might at least fill in
the Balmer cores. The wings are also probably affected by the dynamical
state of the star, leading to poor fits during some phases.
However, during intermediate phases when they closely resemble stable BHB stars,
RR Lyrae variables will not be distinguishable spectroscopically, as borne out by 
the good fits to star 57 in Figure 3. 

We thus sought photometric evidence for variability.
The individual measurements in the
photometric dataset should show variations in magnitude and in
color, which should redden as magnitude declines. However, the 
target field observations span only 18, 40, and 52\mn\ in $U$, $B$, 
and $V$ respectively, much shorter than the RR Lyrae periods of 5 -- 27\hr\ 
\citep{smi95}. Instead we compared the mean 
magnitudes found for all those hot stars observed in both the target field 
and the calibration field obtained the next night, 27 -- 29\,hr later.
Three stars showed definite variability: 57, 124, and 132. All are low-gravity 
stars with \teff\ = 7250\kel, the ones most likely to be RR Lyraes. 
Roughly only one quarter of the variables should be detected 
in a pair of observations made a night apart, due to
improper phasing and the relative constancy of $ab$ variables 
near minimum light.

The stars in our sample most likely to be RR Lyraes from 
the above criteria are 162, 163, 189, 150, 166, 57, 124, 159, 70, 66, and 132.
Overall, we estimate that RR Lyraes 
might comprise as many as a quarter of our low-gravity stars with \teff\ = 
8000 -- 8250\kel, half with 7500 -- 7750\kel, and three-quarters with
\teff\ = 7250\kel.  

\section{Results and Discussion}

Of the 47 stars found to have \teff\ $\geq$ 7250\kel, gravities
\logg\ indicate that nine are possible \popi\ core-hydrogen-burning
stars.  (A tenth, star 163 with the 
unusual gravity \logg\ = 3.5 at \teff\ = 7750\kel, 
is likely to be an RR Lyrae.)
Of the nine, three stars have \teff\ = 7250\kel\ and \logg\ = 4.0, and are
likely to be foreground young stars given their distances. Four clearly
are young \popi\ stars, with \logg\ = 4.0 and \teff\ $<$ 9000\,K. Two
of these show classical peculiarities \citep{jas90}: 91 is a Sr-Cr-Eu
Ap star, with strong \srj\ lines and suitable \bmv\ colors, and 46 has
very weak \caj\ for its Balmer-line and metallic-line strengths, as is
typical of Am stars. Stars 75 and 91 are broad-lined. 
If due to rapid rotation, implied \vsini\ values are 
$\sim$ 110\kms\ and $\sim$ 190\kms\ respectively, values in keeping with 
normal main-sequence A stars (Jaschek \& Jaschek 1990). Two
stars (48 and 182) with \teff\ $\geq$ 9000\kel\ and \logg\ = 3.5 could be either
foreground young stars or \popj\ evolved stars. If the latter, neither
is in the bulge, but rather in front of it.  

Thirty-seven stars are evolved \popj\ core-helium-burning stars on the
HB: those with 8500\kel\ $\geq$ \teff\ $\geq$ 7250\kel\ and 3.0 $\geq$
\logg\ $\geq$ 2.5. As discussed above, among the cooler stars we must
allow for RR Lyraes, which are HB stars but not BHB stars.  We estimate
that two stars with \teff\ $\ge$ 8000\kel, five with 7500 -- 7750\kel,
and seven of the nine with 7250\kel\ should be considered variables.
This leaves the number of bona-fide BHB stars around 23. 

In our sample, no unambiguous BHB stars appear with \teff\ $>$ 8500\kel. 
The Balmer line strengths
reach a maximum near 9500\kel, that of the hottest star we have
discerned, so the possibility exists that we have erroneously assigned
hotter stars to the cooler side. We do not think this has happened,
because in hot stars both the Balmer line profiles and the strengths of
weak atomic lines are well reproduced at the temperatures assigned.

It is premature to conclude that hotter BHB stars are absent from the bulge, however, 
because of the sparse coverage of the spectroscopic dataset at the colors and 
magnitudes appropriate for stars hotter than 10,000\kel\ reddened by \ebmv\ = 
0.3. We may judge this from the \ngc 6791 CMD and the model colors. \cite{cha99} 
match isochrones to the \ngc 6791 CMD, deriving 0.08 $\leq$ \ebmv\ $\leq$ 0.13 
and 13.30 $\leq$ \dmv\ $\leq$ 13.45. Indeed, the colors \citep{mon94,kal95} of 2-17, 
the cool BHB candidate in \ngc6791, are matched to a few hundredths in $U - B$, 
\bmv, and 
\vmi\ by those of the model with \teff\ = 7250\kel, \logg\ = 3.0, and \feh\ = 
+0.3 when \ebmv\ = 0.15 is taken. Adopting \mv\ = 1.1 for this star, for which 
$V$ = 15.0, also reproduces \dmv\ = 13.45. Stars in the bulge would appear 1 mag 
fainter at the same reddening, or 1.5 mag fainter in $V$ if \ebmv\ = 0.3. 
Thus a star with \teff\ = 11,500\kel\ and \logg\ = 4.0 would have $V$ = 16.0 
in \ngc 6791, but would have $V$ = 17.5 in the bulge at \ebmv\ = 0.3, and 
\bmv\ = 0.20 and \umb\ = $-0.06$ based on the models. We have obtained very few 
spectra of stars this faint and this blue (see Figures 1 and 2), 
and so should not be surprised to find none this hot.

The HB stars uncovered by our analysis span a wide metallicity range, $-2.5 \leq
\feh\ \leq\ +0.0$. This is considerably broader than the metallicity
distribution for RR Lyrae stars in Baade's Window at $-4$\dg\ \citep{wt91}, 
which is sharply peaked near $\langle{\rm [Fe/H]}\rangle = -1.0$ and drops 
quickly for $\feh > -0.9$.  
The HB metallicity distribution is also broader than that of 
K giants, but in the opposite sense: the K-giant mean is $\langle\feh\rangle$ = $-0.11$, 
and few K giants are found below $-1.0$ \citep{sad96}.  

The true BHB stars with 8500\kel\ $\geq$ \teff\ $\ge$ 8000\kel\ span a rather 
wide range of distances. Excluding as RR Lyrae stars both star 162 and 
(arbitrarily) star 118 (at 8000\kel), we find for 14 BHB stars a mean distance of 
8.9\kpc\ with an individual deviation $\sigma_i$ = 4.1\kpc. Their radial velocity dispersion 
is 100.5\kms\ with a mean of 22.0\kms. They are metal-poor, 
averaging \feh\ = $-1.59 \pm 0.13$ with $\sigma_i$ = 0.49\,dex. 
The hotter BHB stars in our sample thus are dominated by 
a metal-poor, extended-bulge population reminiscent of the halo. 

The mean metallicity of the cooler HB stars is difficult to judge but is
probably higher. The mean and standard deviation for all 19 low-gravity stars 
with \teff\ $\leq$ 7750\kel\ is \feh\ = $-1.29 \pm 0.18$ and $\sigma_i$ = 0.77. 
For the ten stars remaining after excluding the most likely RR Lyraes listed 
above, we find \feh\ = $-1.10 \pm 0.25$ and $\sigma_i$ = 0.78\,dex. Because of 
the large velocity variability of RR Lyraes, exceeding $100$\kms\ for $ab$ types, 
it is premature to examine the cool HB stars' velocity dispersion.

There are five HB stars with \feh\ $\geq$ $-0.5$, including two with
solar metallicity. Only one has \teff\ $>$ 7500\kel. The five stars are
located an average 8.2\kpc\ away, with a 1$\sigma$ dispersion of
2.9\kpc. Thus they are indeed bulge stars. We cannot exclude the possibility
that both the solar-metallicity stars are RR Lyraes, given their
7500\kel\ temperatures. This is not likely, however, in view of the
paucity of metal-rich $c$ RR Lyrae variables \citep{smi95}, as well as 
the good fits to their spectra and colors. Moreover, the
velocities of these two stars are low, $-56$ and $-45$\kms. Our work
has thus identified solar-metallicity HB stars in the bulge for the
first time.

Coupled with the dichotomy above between the RR Lyrae and K-giant metallicities 
in the bulge, this suggests that the warm and cool BHB stars at this angular distance 
from the Galactic center may be dominated by two different progenitor 
populations. One is metal-poor and the other metal-rich, and their HB-star output 
differs considerably. The metal-poor population, less concentrated to the 
Galactic center, seems to produce many warm and cool BHB stars per K 
giant. The metal-rich population, more concentrated towards the galactic 
center, seems to produce far fewer HB stars per K giant, mostly cool ones. 

Moreover, the metal-poor population might form RR Lyraes in preference to stable 
BHB stars of low temperature, while the metal-rich population might do the opposite. 
This is suggested from the dearth of metal-rich RR Lyraes in both the bulge 
and the solar neighborhood, and by the low temperature of 7250\kel\ \citep{pet98} 
found  for the 
nonvariable metal-rich star 2-17 in \ngc 6791, if a true BHB. Thus 
metal-poor and metal-rich RR Lyraes, like metal-poor and metal-rich sdB's, might 
be formed through a somewhat different assortment of pathways than cool BHB stars. 
Such a possibility is also suggested by measurements of surface rotation \vsini.
Although none of the 27 field RR Lyrae stars measured by \citet{pet96} 
showed \vsini\ $\geq$ 10\kms, over half of the cool and warm BHB stars do so,
in both globular clusters and the field \citep{pet83,pet85,pet95}. Globular-cluster
BHB stars with \teff\ $\geq$ 12,000\kel\ also show very low \vsini\ values
\citep{beh99,beh00}. However, to clarify the pathways of formation of
RR Lyraes and sdB's in metal-rich and metal-poor populations,
we must analyze a much larger sample of bulge HB stars 
in which RR Lyraes are detected photometrically, and 
which goes blue enough and faint enough to detect BHB stars hotter than 12,000\kel. 

Reaching the sdB's -- the hottest, faintest BHB stars -- requires going 
to magnitudes well beyond those of our current survey. 
Given the average magnitude $<V>$ = 18.0 of such stars in \ngc 6791, the discussion above 
suggests they are expected near $V$ = 20.5 at the distance of the bulge when \ebmv\ = 0.3.
At this reddening, the colors of the solar-metallicity Kurucz model with 
\teff\ = 25,000\kel\ and \logg\ = 5.0 are \bmv\ = 0.07 and \umb\ = $-0.68$, 
reaching 0.01 and $-0.86$ at \teff\ = 35,000\kel. 
To detect and count such stars in the future,
we plan to pursue bulge $UBV$ photometry to $V \sim B \sim 21$ and $U \sim 20.5$.
We aim to secure spectroscopy for BHB and blue straggler candidates to $V \sim
18.5$ and perhaps beyond, to characterize their temperatures and metallicities. 
By so extending this survey, we hope to shed light on the
both the mechanism(s) of production of BHB stars and the predisposing factors of the 
UV upturn phenomenon.

\acknowledgements 
We are indebted to S. Allen of U.\ C.\ Santa Cruz and B. Dorman of
Goddard for writing scripts to convert the VAX version of Kurucz codes
into UNIX versions, and thank E. M. Green and J. Liebert for useful discussions.
RCP and DMT gratefully acknowledge support from NSF
grant AST-9900582 to Astrophysical Advances and NSF grants AST-9157038,
INT-9215844, and AST-9820603 to Ohio State University.  Christopher
J.\ Burke assisted with the reduction of the photometry.

\clearpage


\figcaption[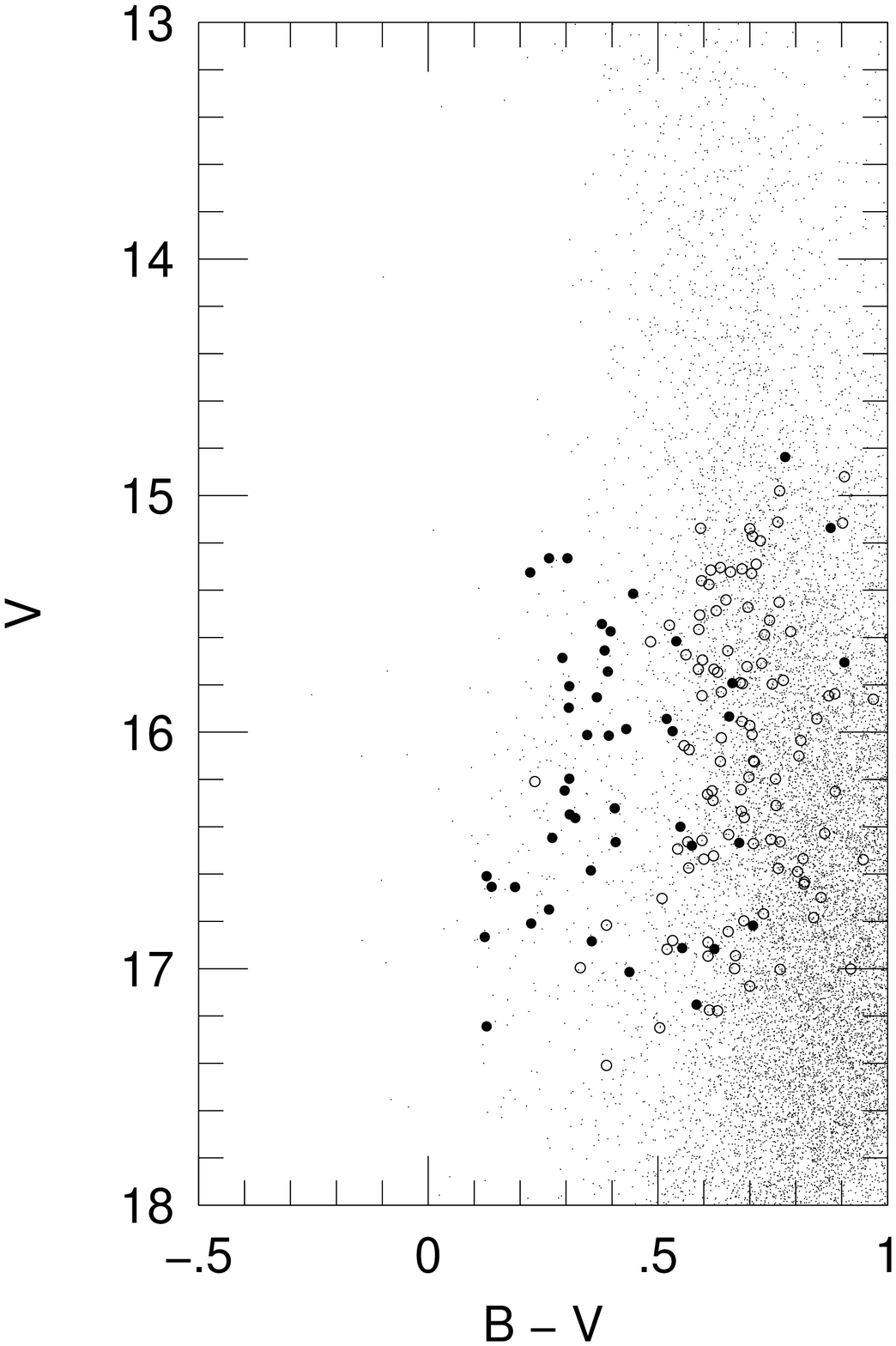]{\label{vbvfig}} Color-magnitude diagram in
$V$ vs.\ \bmv\ for the field studied here.  
The large symbols represent stars observed spectroscopically,
with filled points for the hot stars listed
in Table 1, and open circles for the others.
Small points are plotted for 17\% of all other stars with $V$ and $B$ photometry,
to show the relationship of the stars observed spectroscopically. 
Main-sequence contamination becomes significant redward of \bmv\ = 0.5.

\figcaption[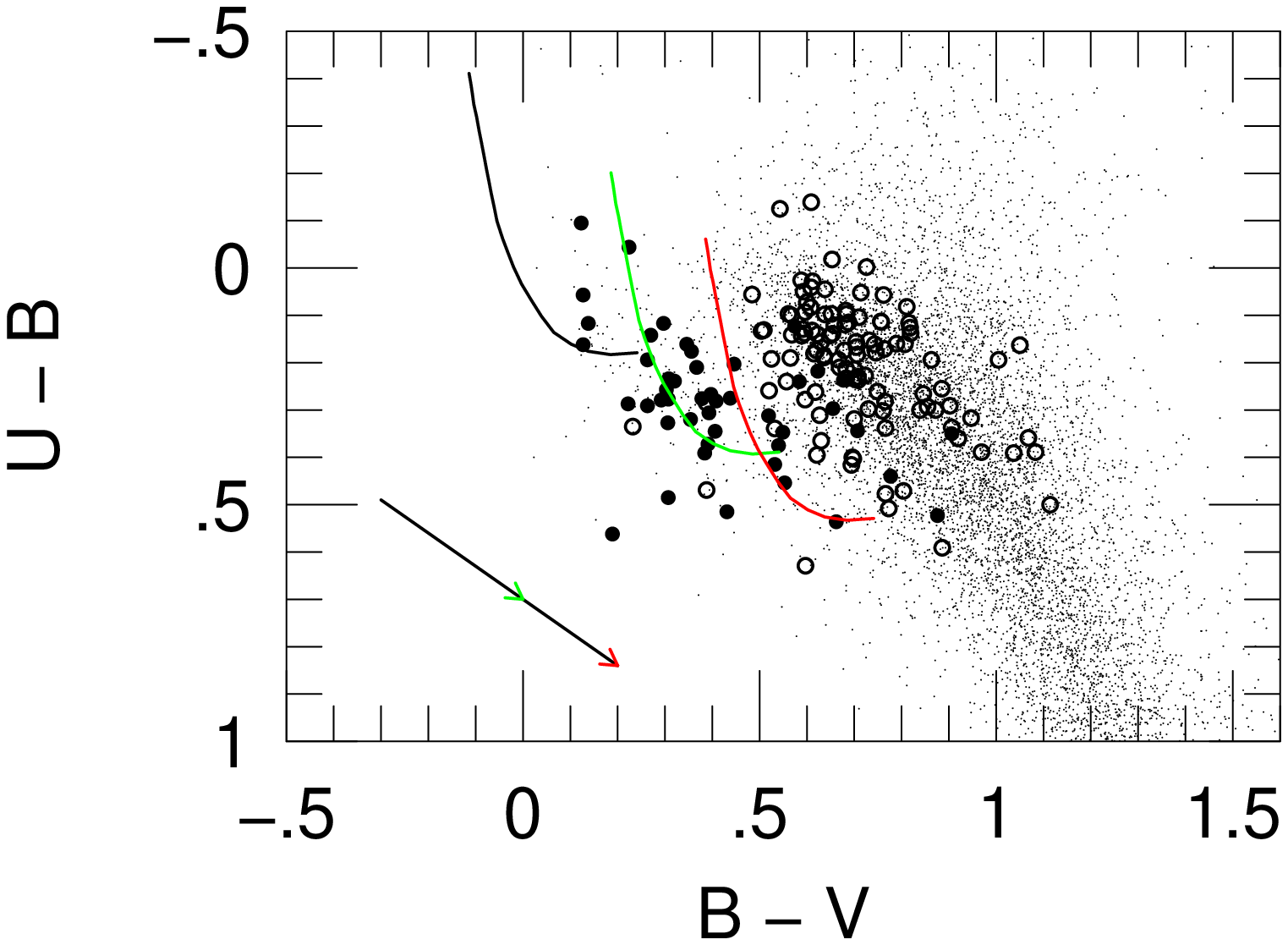]{\label{ubbvfig}} Color-color diagram in $UBV$.
The large symbols are as described for the previous figure. 
Small points are plotted for 17\% of all other stars with $U$, $B$, and $V$ photometry.
The concentration
of points corresponds to the main-sequence turnoff
of the foreground disk. 
The curves represent the colors of BHB stars
with \feh\ = 0 from \teff, \logg\ = 7250\kel, 2.7 to 13,000\kel, 4.1, reddened by 
\ebmv\ = 0.0, 0.3, (the mean for the field), and 0.5. The
reddening vector $E(U-B)$ = $0.70 \times \ebmv$ of \citet{kur91} 
is depicted by the arrow, with heads at the latter reddenings. 
Except where  \teff\ $\leq$ 7500\kel, the BHB model $UBV$ colors at a
particular \teff\ and \logg\
change by $<$ 0.01 mag in going to
\feh\ = +0.3 or $-0.5$.
This illustrates that the colors of BHB stars, regardless of metallicity,
remain distinguishable from the colors
of main-sequence stars cooler than 7000\kel\ with near-solar metallicity until
\ebmv\ $>$ 0.50. 

\figcaption[fig3a-d.ps]{\label{specfig}} Observed spectra (heavy line)
are compared to calculations (light lines) for six stars of various
temperatures and metallicities. The normalized spectral flux is plotted
against wavelength in \angs, with successive spectra displaced upwards
by 20\% of the continuum. 
The Balmer lines \Hdelta\ (4101\angs), \Hgamma\ (4340\,\AA), and
\Hbeta\ (4861\angs) are the strongest features
shown in the spectrum. Many weak metal lines are seen superimposed
on the Balmer line wings, especially at lower temperatures, higher 
metallicities, and bluer wavelengths. 
At the upper right of each spectrum, the 
model atmosphere parameters \teff, \logg, \feh, and \alfe\ used for 
the best-fit calculation are listed
next to the number of each star. 
(a) 4060 -- 4230\,\AA. 
(b) 4230 -- 4400\angs.  (c) 4400 -- 4570\angs. (d) 4780 -- 4950\angs.

\figcaption[fig4.ps]{\label{specfig4}} As in Figure 3, 
but adopting models that differ by 0.5 dex in \logg. (a) 
4230 -- 4400\angs.  (b) 4780 -- 4950\angs.

\clearpage

\begin{deluxetable}{rrrrrrrrrrrrrrrrrrr}
\tablenum{1}
\tablecolumns{19}
\tabletypesize{\scriptsize}
\tablecaption{\bf Astrometry, Photometry, and Stellar Parameters for Hot Stars}
\tablewidth{0pt}
\rotate
\tablehead{
\colhead{\#} & 
\colhead{ID} & 
\multicolumn{2}{c}{RA~~~(2000)~~~Dec} &
\colhead{$V$} & 
\colhead{$\sigma$} & 
\colhead{\bmv} & 
\colhead{$\sigma$} & 
\colhead{\umb} & 
\colhead{$\sigma$} & 
\colhead{V$_{\rm rad}$} & 
\colhead{\teff} & 
\colhead{\logg} & 
\colhead{$Z$} & 
\colhead{D} & 
\colhead{\mv} & 
\colhead{$E$} & 
\colhead{$\delta_U$} & 
\colhead{$\delta_B$}
}

\startdata

48 & 64467 & 18 06 35.13 & -34 40 08.71 & 16.02 & 0.09 & 0.39 & 0.06 & 0.31 & 0.05 & 96.1 & 9500 & 3.5 & -0.5 & 5.1 & 1.1 & 0.45 & 0.03 & -0.01\\
182 & 730 & 18 08 55.89 & -35 31 20.54 & 15.94 & 0.08 & 0.52 & 0.05 & 0.31 & 0.23 & -107.6 & 9000 & 3.5 & -0.5 & 3.8 & 1.5 & 0.50 & -0.07 & 0.03\\
168 & 101519 & 18 05 19.18 & -35 36 41.02 & 15.90 & 0.07 & 0.31 & 0.04 & 0.33 & 0.02 & 85.0 & 8500 & 3.0 & -2.5 & 7.6 & 0.4 & 0.35 & 0.01 & -0.05\\
69 & 102439 & 18 05 15.84 & -34 41 54.97 & 15.42 & 0.08 & 0.45 & 0.05 & 0.20 & 0.03 & -49.5 & 8500 & 3.0 & -2.0 & 5.8 & 0.5 & 0.35 & -0.11 & 0.09\\
135 & 126450 & 18 04 27.09 & -35 06 55.12 & 17.24 & 0.10 & 0.13 & 0.06 & 0.16 & 0.05 & 5.9 & 8500 & 3.0 & -2.0 & 19.4 & 0.5 & 0.10 & 0.01 & 0.01\\
185 & 124221 & 18 04 32.03 & -35 25 42.69 & 15.54 & 0.05 & 0.38 & 0.03 & 0.28 & 0.03 & -118.1 & 8500 & 3.0 & -2.0 & 6.2 & 0.5 & 0.35 & -0.04 & 0.02\\
20 & 35832 & 18 07 37.56 & -34 56 19.48 & 16.25 & 0.09 & 0.30 & 0.05 & 0.12 & 0.03 & -195.7 & 8500 & 3.0 & -1.5 & 10.1 & 0.6 & 0.20 & -0.10 & 0.08\\
36 & 42402 & 18 07 23.70 & -35 03 24.99 & 15.99 & 0.07 & 0.43 & 0.05 & 0.52 & 0.08 & -97.9 & 8500 & 3.0 & -1.5 & 5.9 & 0.6 & 0.50 & 0.10 & -0.07\\
126 & 137564 & 18 04 04.55 & -34 57 06.47 & 15.33 & 0.07 & 0.22 & 0.05 & 0.29 & 0.04 & 27.7 & 8500 & 3.0 & -1.5 & 6.2 & 0.6 & 0.25 & 0.04 & -0.04\\
151 & 150654 & 18 03 39.04 & -35 09 32.24 & 16.47 & 0.10 & 0.41 & 0.06 & 0.28 & 0.04 & 78.4 & 8500 & 3.0 & -1.5 & 9.0 & 0.6 & 0.35 & -0.04 & 0.05\\
160 & 93003 & 18 05 37.31 & -35 44 00.07 & 16.35 & 0.04 & 0.31 & 0.04 & 0.23 & 0.07 & 123.7 & 8500 & 3.0 & -1.5 & 9.2 & 0.6 & 0.30 & -0.05 & 0.00\\
187 & 62992 & 18 06 39.57 & -35 09 41.24 & 16.00 & 0.12 & 0.53 & 0.07 & 0.42 & 0.17 & 101.3 & 8500 & 3.0 & -1.5 & 5.9 & 0.6 & 0.50 & 0.00 & 0.03\\
139 & 150568 & 18 03 39.22 & -35 14 12.89 & 16.75 & 0.08 & 0.26 & 0.05 & 0.19 & 0.05 & 190.9 & 8500 & 3.0 & -1.0 & 12.2 & 0.7 & 0.20 & -0.03 & 0.05\\
162 & 97031 & 18 05 28.70 & -35 38 43.41 & 16.66 & 0.17 & 0.19 & 0.11 & 0.56 & 0.05 & 120.0 & 8500 & 3.0 & -1.0 & 8.8 & 0.7 & 0.40 & 0.21 & -0.22\\
7 & 32360 & 18 07 45.39 & -35 02 34.49 & 15.27 & 0.02 & 0.30 & 0.02 & 0.26 & 0.02 & 102.8 & 8500 & 3.0 & -0.5 & 5.1 & 0.8 & 0.30 & -0.03 & -0.01\\
75 & 109811 & 18 05 00.96 & -34 53 37.19 & 15.27 & 0.06 & 0.26 & 0.04 & 0.29 & 0.03 & -43.3 & 8250 & 4.0 & 0.3 & 3.5 & 2.1 & 0.15 & 0.07 & -0.03\\
91 & 127739 & 18 04 24.35 & -34 57 03.42 & 15.14 & 0.13 & 0.88 & 0.08 & 0.52 & 0.05 & -31.6 & 8250 & 4.0 & 0.0 & & \multicolumn{3}{c}{Sr-Cr-Eu Ap star}\\
110 & 103837 & 18 05 13.93 & -35 19 24.65 & 15.69 & 0.05 & 0.29 & 0.03 & 0.28 & 0.03 & 87.7 & 8250 & 3.0 & -1.5 & 7.3 & 0.6 & 0.25 & 0.00 & 0.01\\
118 & 108764 & 18 05 03.20 & -34 57 39.02 & 16.59 & 0.06 & 0.35 & 0.05 & 0.32 & 0.06 & -40.3 & 8250 & 3.0 & -1.5 & 10.3 & 0.6 & 0.30 & 0.01 & 0.02\\
152 & 139664 & 18 04 00.84 & -35 32 38.68 & 16.65 & 0.09 & 0.14 & 0.06 & 0.12 & 0.06 & 19.1 & 8250 & 2.5 & -2.0 & 14.8 & 0.5 & 0.10 & -0.04 & 0.02\\
46 & 92029 & 18 05 38.21 & -35 07 07.84 & 15.85 & 0.05 & 0.37 & 0.04 & 0.21 & 0.07 & 34.8 & 8000 & 4.0 & 0.3 & 4.1 & 2.3 & 0.15 & -0.01 & 0.03\\
186 & 11492 & 18 08 32.69 & -35 30 51.35 & 16.88 & 0.09 & 0.36 & 0.06 & 0.18 & 0.07 & 55.9 & 8000 & 4.0 & 0.0 & 6.7 & 2.3 & 0.15 & -0.02 & 0.03\\
188 & 74233 & 18 06 16.53 & -35 32 03.66 & 17.01 & 0.06 & 0.44 & 0.06 & 0.28 & 0.09 & -13.4 & 8000 & 3.0 & -1.0 & 11.1 & 0.7 & 0.35 & -0.09 & 0.03\\
195 & 9137 & 18 08 37.43 & -35 24 12.32 & 16.61 & 0.05 & 0.13 & 0.04 & 0.06 & 0.17 & 4.5 & 8000 & 2.5 & -2.0 & 15.5 & 0.5 & 0.05 & -0.09 & 0.04\\
163 & 85527 & 18 05 52.69 & -35 38 07.62 & 17.15 & 0.07 & 0.58 & 0.05 & 0.24 & 0.11 & -204.1 & 7750 & 3.5 & 0.0 & 7.8 & 1.6 & 0.35 & -0.14 & 0.07\\
52 & 69661 & 18 06 25.17 & -35 04 17.66 & 16.36 & 0.05 & 0.32 & 0.03 & 0.24 & 0.05 & -200.1 & 7750 & 3.0 & -1.5 & 10.7 & 0.6 & 0.20 & -0.01 & 0.01\\
105 & 125288 & 18 04 29.80 & -35 25 40.44 & 14.84 & 0.17 & 0.78 & 0.10 & 0.44 & 0.03 & 68.2 & 7750 & 3.0 & -1.5 & 3.2 & 0.6 & 0.55 & -0.05 & 0.14\\
179 & 68186 & 18 06 28.84 & -35 15 47.36 & 15.74 & 0.03 & 0.39 & 0.02 & 0.37 & 0.04 & 106.0 & 7750 & 3.0 & -1.5 & 7.0 & 0.6 & 0.30 & 0.05 & -0.01\\
54 & 90565 & 18 05 40.56 & -34 45 46.90 & 16.45 & 0.05 & 0.27 & 0.04 & 0.14 & 0.05 & -156.6 & 7750 & 2.5 & -2.5 & 13.1 & 0.4 & 0.15 & -0.08 & 0.05\\
147 & 151052 & 18 03 38.34 & -35 29 41.55 & 15.93 & 0.05 & 0.66 & 0.04 & 0.30 & 0.03 & 30.4 & 7500 & 3.0 & -0.5 & 6.5 & 0.8 & 0.35 & -0.08 & 0.14\\
76 & 139386 & 18 04 00.82 & -34 40 19.26 & 16.91 & 0.11 & 0.55 & 0.07 & 0.45 & 0.10 & -45.9 & 7500 & 3.0 & 0.0 & 9.7 & 0.9 & 0.35 & 0.04 & 0.02\\
98 & 94051 & 18 05 33.58 & -34 51 35.38 & 16.40 & 0.14 & 0.55 & 0.09 & 0.35 & 0.05 & -34.9 & 7500 & 3.0 & 0.0 & 8.2 & 0.9 & 0.30 & -0.03 & 0.07\\
189 & 110738 & 18 04 59.68 & -35 23 19.23 & 16.87 & 0.08 & 0.12 & 0.05 & -0.10 & 0.06 & -102.1 & 7500 & 2.5 & -2.5 & 19.6 & 0.4 & 0.00 & -0.22 & 0.01\\
190 & 7372 & 18 08 42.14 & -35 38 09.95 & 15.81 & 0.08 & 0.31 & 0.05 & 0.28 & 0.05 & -100.9 & 7500 & 2.5 & -1.5 & 8.9 & 0.6 & 0.15 & 0.04 & 0.04\\
150 & 146951 & 18 03 46.38 & -35 27 40.48 & 16.92 & 0.06 & 0.62 & 0.04 & 0.22 & 0.12 & 27.6 & 7500 & 2.5 & -1.0 & 10.6 & 0.7 & 0.35 & -0.16 & 0.16\\
140 & 137700 & 18 04 04.81 & -35 38 48.13 & 16.47 & 0.07 & 0.68 & 0.06 & 0.24 & 0.23 & -3.5 & 7250 & 4.0 & 0.3 & 2.1 & 3.9 & 0.30 & -0.01 & 0.06\\
77 & 138795 & 18 04 02.08 & -34 45 35.85 & 15.62 & 0.06 & 0.54 & 0.04 & 0.38 & 0.05 & -180.7 & 7250 & 4.0 & -2.0 & 1.2 & 3.9 & 0.45 & 0.18 & -0.16\\
80 & 107827 & 18 05 04.77 & -34 46 41.88 & 16.82 & 0.13 & 0.71 & 0.08 & 0.34 & 0.11 & -146.4 & 7250 & 4.0 & 0.0 & 2.0 & 3.9 & 0.45 & 0.02 & -0.03\\
166 & 103531 & 18 05 15.17 & -35 44 03.59 & 16.97 & 0.08 & & & & & 78.7 & 7250 & 3.0 & -2.0 & 0.5\\
57 & 67728 & 18 06 29.21 & -35 02 17.15 & 15.79 & 0.07 & 0.66 & 0.06 & 0.54 & 0.08 & -78.3 & 7250 & 3.0 & -1.0 & 4.8 & 0.7 & 0.55 & 0.07 & -0.07\\
192 & 8375 & 18 08 38.12 & -35 10 02.29 & 15.57 & 0.07 & 0.40 & 0.06 & 0.27 & 0.06 & 119.1 & 7250 & 3.0 & -1.0 & 7.1 & 0.7 & 0.20 & 0.05 & -0.01\\
124 & 101879 & 18 05 17.72 & -35 09 45.82 & 15.66 & 0.05 & 0.38 & 0.03 & 0.39 & 0.07 & 313.6 & 7250 & 3.0 & -0.5 & 6.5 & 0.8 & 0.25 & 0.10 & -0.08\\
159 & 114296 & 18 04 52.89 & -35 44 47.05 & 16.81 & 0.13 & 0.22 & 0.08 & -0.04 & 0.05 & 47.1 & 7250 & 2.5 & -2.5 & 19.1 & 0.4 & 0.00 & -0.14 & 0.06\\
70 & 114182 & 18 04 51.78 & -34 40 08.52 & 16.20 & 0.27 & 0.31 & 0.17 & 0.49 & 0.08 & 94.2 & 7250 & 2.5 & -2.0 & 8.2 & 0.4 & 0.40 & 0.11 & -0.24\\
66 & 101703 & 18 05 17.27 & -34 40 15.12 & 16.48 & 0.07 & 0.57 & 0.05 & 0.12 & 0.06 & -196.9 & 7250 & 2.5 & -1.0 & 11.0 & 0.5 & 0.25 & -0.18 & 0.15\\
197 & 26411 & 18 07 59.16 & -35 15 03.12 & 16.01 & 0.05 & 0.35 & 0.04 & 0.16 & 0.05 & 159.3 & 7250 & 2.5 & -1.0 & 10.0 & 0.7 & 0.10 & -0.04 & 0.07\\
132 & 157356 & 18 03 26.02 & -35 36 33.84 & 16.32 & 0.05 & 0.41 & 0.04 & 0.35 & 0.04 & -157.7 & 7250 & 2.5 & -1.0 & 9.3 & 0.7 & 0.25 & 0.05 & -0.02\\

\enddata

\tablecomments{Units: RA in h, m, s; Dec in~\dg,$^\prime$,$^{\prime \prime}$; \teff\ in~\kel; $V_{\rm rad}$ in~\kms\ relative to the internal F-star template; D in kpc.}

\end{deluxetable}

\clearpage

\end{document}